

\def\IR{{\hbox{{\rm I}\kern-.2em\hbox{\rm R}}}}
\def\IB{{\hbox{{\rm I}\kern-.2em\hbox{\rm B}}}}
\def\IC{{\ \hbox{{\rm I}\kern-.6em\hbox{\bf C}}}}

\def\IZ{{\hbox{{\rm Z}\kern-.4em\hbox{\rm Z}}}}
\def\to{\rightarrow}
\def\underarrow#1{\vbox{\ialign{##\crcr$\hfil\displaystyle
{#1}\hfil$\crcr\noalign{\kern1pt
\nointerlineskip}$\longrightarrow$\crcr}}}
%

\input phyzzx
\overfullrule=0pt
\tolerance=5000
\overfullrule=0pt
\twelvepoint

\twelvepoint
\pubnum{IASSNS-HEP-91/51}
\date{August, 1991}
\titlepage
\title{GROUND RING OF TWO DIMENSIONAL STRING THEORY}
\vglue-.25in
\author{Edward Witten\foot{Research supported in part by NSF Grant
PHY86-20266.}}
\medskip
\address{School of Natural Sciences
\break Institute for Advanced Study
\break Olden Lane
\break Princeton, NJ 08540}
\bigskip
\abstract{String theories with two dimensional space-time target spaces
are characterized by the existence of a ``ground ring'' of operators
of spin $(0,0)$.  By understanding this ring, one can understand
the symmetries of the theory and illuminate the relation of the critical
string theory to matrix models.  The symmetry groups that arise are, roughly,
the area preserving diffeomorphisms of a two dimensional phase space that
preserve the fermi surface (of the matrix model) and the
volume preserving diffeomorphisms of a three dimensional cone.
The three dimensions in question are the matrix eigenvalue, its canonical
momentum, and the time of the matrix model. }
\endpage
\REF\segal{J. Goldstone, unpublished; V. G. Kac, ``Contravariant Form
For Infinite Dimensional Lie Algebras and Superalgebras,'' in
{\it Group Theoretical Methods in Physics}, ed. W. Beiglbock et. al.,
Lecture Notes in Physics,
vol. 94 (Springer-Verlag, 1979); G. Segal, ``Unitary Representations Of Some
Infinite Dimensional Groups,'' Comm. Math. Phys. {\bf 80} (1981) 301.  }
\REF\gross{D. J. Gross, I. R. Klebanov, and M. J. Newman, ``The Two-Point
Correlation Function Of The One Dimensional Matrix Model,'' Nucl. Phys.
{\bf B350} (1991) 621; D. J. Gross and I. R. Klebanov, PUPT-1241 (1991).}
\REF\polyakov{A. M. Polyakov, ``Self-Tuning Fields And Resonant
Correlations in $2D$ Gravity,'' Mod. Phys. Lett. {\bf A6} (1991) 635.}
\REF\witten{E. Witten, ``On String Theory And Black Holes,''
Phys. Rev. {\bf D44} (1991) 314.}
\REF\ellis{J. Ellis, N. E. Mavromatos, and D. V. Nanopoulos,
``Quantum Coherence And Two-Dimensional Black Holes,''
CERN-TH.6147/91, CTP-TAMU-41/91.}
\REF\grossdan{D. J. Gross and U. H. Danielson, ``On The Correlation
Functions Of The Special Operators in $c=1$ Quantum Gravity,''
PUPT-1258 (1991).}
\REF\moore{G. Moore and N. Seiberg, ``From Loops To Fields In Two
Dimensional Gravity,'' Rutgers preprint.}
\REF\wadia{G. Mandal and S. Wadia, private communication and to appear.}
\chapter{Introduction}

Critical string theory with two dimensional target space time
has been the subject of much recent interest.  The simplest target
space is a flat space-time with a linear dilaton field, and perhaps
an additional ``Liouville'' term.  The
corresponding world-sheet Lagrangian is
$$\eqalign{ L={1\over 4\pi\alpha'}&\int d^2x \sqrt
h\left(h^{ij}\partial_iX\partial_jX
+h^{ij}\partial_i\phi\partial_j\phi\right)-{1\over 2\pi\sqrt \alpha'}\int
d^2x \sqrt h \cdot \phi R^{(2)}\cr &
+\mu\int d^2x\sqrt h\,\,\,\exp(-2\phi/\sqrt
{\alpha'}).  \cr} \eqn\simplest$$
($h$ is the world-sheet metric, $R^{(2)}$ is the Ricci scalar, and $\mu$ is
the cosmological constant.)
The ``matter'' field $X$ is decoupled from the ``Liouville'' field
$\phi$ (and from the ghosts that appear upon quantization), so some
vertex operators can be constructed as products of $X$ operators and
$\phi$ operators.  The obvious conformal fields
constructed from $X$ are standard vertex operators $e^{ipX}$.
These depend on the continuously variable parameter $p$.
In addition, at particular values of the momentum, additional
``discrete'' primary fields appear.  The first of these, at zero
momentum, is the current $\partial X$ (or, when left and right movers
are combined, the gravitional vertex operator $\bar\partial X\partial X$).
These discrete states have long been known [\segal].
In the context of two dimensional string theory they were discussed
at an early stage by
N. Seiberg (unpublished) and have showed up
in matrix model calculations [\gross]. It has been argued that they
should be interpreted in terms of a ``stringy'' target space topological
field theory [\polyakov]; this indicates that understanding
these modes is essential for
extracting the real physical lessons of two dimensional string theory.
These modes may be important in understanding
two dimensional black holes [\witten,\ellis].
They have been the subject of several recent studies [\grossdan,\moore,\wadia].

In addition to the discrete vertex operators that have usually
been considered, which occur (if one considers $(1,1)$ operators)
at ghost number zero,
there are also discrete vertex operators which occur at the same
values of the momenta at adjacent values of the ghost number.
This is obvious in the BRST formalism.  If one has a family of states
$|\alpha(p)\rangle$, of unit norm,
parametrized by a momentum variable $p$,
and BRST invariant only at, say, $p=p_0$, then one has
$$Q|\alpha(p)\rangle =f(p)|\beta(p)\rangle, \eqn\impo$$
where $|\beta(p)\rangle$ is of unit norm and $f(p)$ vanishes
precisely at $p=p_0$.  Then $|\alpha(p_0)\rangle$ and $|\beta(p_0)\rangle$
are a pair of discrete states at adjacent values of the ghost number.

\REF\zuckerman{B. Lian and G. Zuckerman, ``New Selection Rules And
Physical States in 2D Gravity,'' Phys. Lett. {\bf B254} (1991) 417,
``2D Gravity With $c=1$ Matter,''
Yale University preprint
(1991).}
The whole pattern of discrete states of various ghost number has
been thoroughly described by Lian and Zuckerman [\zuckerman].  However,
the physical
consequences of the existence of discrete states of non-standard ghost
number have not yet been considered.  This is the problem that
we will consider in the present paper.

\chapter{The Discrete States}

To begin with, we consider the holomorphic part of the current algebra.
Thus, $X$ is a free field with
$$ X(z)X(w)\sim -\ln(z-w). \eqn\funko$$
The stress tensor is $T_{zz}=-(\partial_zX)^2/2$.
The tachyon operators
$V_p=e^{ipX}$ are conformal fields of dimension $p^2/2$.

For reasons that are standard and
will become clear, it is convenient to first consider
the theory compactified at the $SU(2)$ radius and with zero cosmological
constant.
At the $SU(2)$  radius, the allowed values of the momenta are
$p=n/\sqrt 2$, with $n\in \IZ$.  There is an  $SU(2)$ current algebra
with $(T_+,T_3,T_-)$ represented by $(e^{iX\sqrt 2},i\sqrt 2
\partial X,e^{-iX\sqrt 2})$.
For $s\geq 0$,there is no singularity in the short distance product
$$ e^{iX\sqrt 2}(z)\cdot e^{isX/\sqrt 2}(w) \eqn\unko$$
for $z\to w$, so the operator $e^{isX/\sqrt 2}$ transforms as the highest
weight of an $SU(2)$ representation.  In view of the $J_3$ value, this
is a representation of spin $s$.  By repeatedly acting with the lowering
operators $J_-$, we fill out the $SU(2)$ multiplets with operators
$V_{s,n}$, $n\in \{s,s-1,\dots,-s\}$, such that $V_{s,s}=e^{isX/\sqrt 2},$
$V_{s,-s}=e^{-isX/\sqrt 2}$.\foot{The latter holds since a rotation about
the $J_1$ axis gives $X\to -X$ and reverses the sign of $J_3$.}
The $V_{s,n}$ are all Virasoro primaries, since the $V_{s,s}$ are standard
primary fields and the $SU(2)$ current algebra intertwines with Virasoro.
For $|n|=s$, the $V_{s,n}$ are standard tachyon operators at particular
momenta, but for $|n|<s$, they are new objects, ``discrete primaries.''
The conformal dimension of $V_{s,n}$ is $s^2/4$.
More detailed investigation [\segal,\zuckerman] shows that regardless of
the radius of compactification of the $X$ field, the only primary
fields of the $X$ system are the tachyon vertex operators and the
$V_{s,n}$.  At an arbitrary radius, the discrete primaries are whatever
combinations of $V_{s,n}$'s with anti-holomorphic counterparts are
permitted (by the standard momentum restrictions) at that particular radius.

\subsection{Coupling To Other Degrees Of Freedom}

Now, let us couple the $V_{s,n}$'s to the
rest of the system.
First, we consider the Liouville field $\phi$.
Its stress tensor is $T_{zz}=-(1/2)(\partial \phi)^2+\sqrt 2\phi$.
The operator $e^{iw\phi}$ has conformal dimension $w^2/2+iw\sqrt 2$.
Looking for a spin one primary field of the form $V_{s,n}e^{iw\phi}$,
we see that there are two possibilities:
$$W_{s,n}^\pm=V_{s,n} \cdot e^{\sqrt 2\phi \mp s\phi\sqrt 2}.\eqn\rinko$$
Note that $W_{s,n}^\pm$ has momentum $(n,i(-1\pm s))\cdot\sqrt 2$.
The importance of the spin one condition is that $(1,1)$ vertex operators
corresponding to infinitesimal deformations of the theory can be
constructed in the form
$$ W_{s,n}^\pm(z)\cdot \overline W_{s,n'}^{\pm}(z)\eqn\onko$$
where $\overline W_{s,n}$ are the antiholomorphic operators analogous
to the $W_{s,n}$.  Notice that as the Liouville field is noncompact,
the Liouville momenta must be matched between left and right movers,
which is the reason that the left-movers and right-movers in
\onko\ share the same value of $s$ and of the sign $\pm$.
At the $SU(2)$ radius, there is no similar matching of $n$.
Note that $W_{0,0}^+=W_{0,0}^-$ is the cosmological constant operator
$e^{\phi\sqrt 2}$.

Now we wish to include the ghosts, which are fermi fields $b$ and $c$
of spins 2 and $-1$, with $T_{zz}=-2b_{zz}\partial_zc^z-\partial_zb_{zz}
\cdot c^z$.
One of the important merits of the ghosts is that in their presence, the
physical state condition can be incorporated by considering the cohomology
of the BRST operator $Q$.
In the standard fashion, the spin one fields $W_{s,n}^{\pm}$
are related to spin 0, BRST invariant fields
$$ Y_{s,n}^{\pm}=cW_{s,n}^{\pm}     \eqn\offgo$$
of ghost number 1.

\section{Additional States At Other Values Of The Ghost Number}

\def\co{{\cal O}}
This cannot be the whole story, for a reason indicated in the introduction.
The $Y_{s,n}^{\pm}$ of $|n|=s$
are not truly discrete (but are special tachyon states).  However,
the ghost number one discrete operators
$Y_{s,n}^{\pm}$ of $|n|<s$ must have ``partners''
at an adjoining value of the ghost number, namely
0 or 2.  In fact, the $Y_{s,n}^{+}$ have partners at ghost number 0,
and the $Y_{s,n}^{-}$ have partners at ghost number 2 [\zuckerman].
The ghost number 0 partner of $Y_{s,n}^+$ will be called in this
paper $\co_{u,n}$, with $u=s-1$.  The momentum of $\co_{u,n}$ is thus
$(n,iu)\cdot \sqrt 2$.
Since $|n|<s$, we have $|n|\leq u$,
so the allowed values of $u$ and $n$ are $u=0,1/2,1,\dots$
and $n=u,u-1,\dots,-u$.

For instance, the first case of a $Y_{s,n}^+ $ with $|n|<s$ is
$Y_{1,0}^+=c\partial X$, a discrete
primary field of ghost number 1 and
momentum $(0,0)$.  The corresponding
discrete primary field of ghost number 0 and momentum $(0,0)$ is the
identity operator, $\co_{0,0}=1$.
At the next level, there are two discrete primaries, $Y_{3/2,1/2}^+$
and $Y_{3/2,-1/2}^+$, of spin 0 and ghost number 1.  The corresponding
spin 0, ghost number 0, operators are
$$\eqalign{ x & = \co_{1/2,1/2}=\left(cb+{i\over \sqrt 2}(\partial X
-i\partial \phi)\right)\cdot e^{i(X+i\phi)/\sqrt 2}  \cr
y & = \co_{1/2,-1/2}=\left(cb-{i\over \sqrt 2}(\partial X
+i\partial \phi)\right)\cdot e^{-i(X-i\phi)/\sqrt 2}.  \cr} \eqn\tigof$$
Some details about their construction can be found in appendix 1.

\section{The Ground Ring}

There are two immediate
reasons that the existence of spin zero, ghost number
zero operators is of vital importance:

(1) It leads to the existence of symmetries of the conformal field
theory under study.

(2) It leads to the existence of a ring, which we will call the
``ground ring,'' which largely controls the properties of the theory,
as we will see.

To understand the first point, note that we already know of the existence
of infinitely many spin $(1,0)$ and $(0,1)$ operators of zero ghost
number, namely $W_{s,n}^\pm$ and $\overline W_{s,n}^\pm$.
Because of the Liouville momentum that they carry, which must be
matched between left and right movers, to construct true quantum field
operators as opposed to the holomorphic or antiholomorphic chiral vertex
operators of the above discussion, $W$ (or $\overline W$)
should not be considered by itself but must be paired with
antiholomorphic (or holomorphic) fields of the same Liouville momentum.
One way to do this is to pair $W\cdot\overline W$, as in \onko,
to make fields of spin $(1,1)$, corresponding to infinitesimal moduli.
To make quantum field
operators of spin $(1,0)$, however, we must combine
a $W$ with an antiholomorphic spin zero field of the same Liouville
momentum, and conversely for $\overline W$.
The spin $(1,0)$ and spin $(0,1)$ currents of the theory are consequently
$$\eqalign{J_{s,n,n'} & = W_{s,n}^+\overline\co_{s-1,n'} \cr
           \overline J_{s,n,n'} & = \co_{s-1,n}\overline W_{s,n'}^+ \cr}
                       \eqn\ruffox$$
On general grounds, these currents generate a Lie algebra of symmetries
-- which we will presently determined to be the Lie algebra of volume
preserving diffeomorphisms of a certain three manifold.

The spin 0 BRST invariant operators generate a commutative, associative
ring for the following reason.  Let ${\cal O}$ and ${\cal O}'$ be two
such operators.  The operator product ${\cal O}(z)\co'(0)$ is BRST invariant,
so all the terms in its short distance expansion for $z\to 0$ are BRST
invariant.  Negative powers of $z$ may arise in this short distance expansion,
but the operators multiplying the negative powers of $z$ are negative
dimension operators which must be BRST commutators (as there is no BRST
cohomology at negative dimension in this theory).  Therefore, modulo the
BRST commutators, the short distance limit of ${\cal O}(z){\cal O}'(0)$
is some BRST invariant spin zero operator ${\cal O}''(0)$ (which may vanish):
$${\cal O}(z){\cal O}'(0)\sim \co''(0)+\{Q,\dots\}. \eqn\turfod$$
This is the desired multiplication law, $\co\cdot \co'=\co''$.
This procedure obviously defines a commutative, associative ring.

By combining left and right movers in the usual
way we can form spin $(0,0)$ quantum field operators
$${\cal V}_{u,n,n'}=\co_{u,n}\overline\co_{u,n'} \eqn\urfod$$
which can be multiplied in just the same way.  We will call the
ring of the ${\cal V}$'s the ground ring of the theory, and we will call
the ring of the chiral ${\cal O}$'s the chiral ground ring.

Since the ground ring is naturally associated with the conformal field
theory, any symmetries of the theory must be symmetries of the ground
ring.  This will make it easy to determine the symmetry algebra.
In fact, if $J$ is a spin one current generating a symmetry, then the
action of $J$ on a spin $(0,0)$ operator $\co$ is defined by
$$J(\co(P))={1\over 2\pi i}\oint_{{\cal C}} J(z)\cdot \co(P)       \eqn\ufgo$$
(modulo BRST commutators)
where ${\cal C}$ is any contour surrounding the point $P$.
Standard contour manipulations show that  if $J$ and $J'$ are two
currents and $[J,J']$ is their commutator then
$$[J,J'](\co(P))=J(J'(\co(P)))-J'(J(\co(P))),  \eqn\fgo$$
so that the $\co(P)$'s form a representation of the symmetry algebra.
What is more unusual is that
since a contour enclosing two points $P$ and $Q$ is homologous
to the sum of a contour surrounding $P$ and a contour surrounding $Q$, we get
$$J(\co(P)\co'(Q))=J(\co(P))\cdot \co(Q)+\co(P)\cdot J(\co'(Q)),\eqn\ifgo$$
which in the limit $P\to Q$ gives
$$J(\co\cdot \co')=J(\co)\cdot \co'+\co\cdot J(\co'). \eqn\pofgo$$
Thus, the $J$'s act as derivations of the ring of $\co$'s.
These considerations can be applied in two different ways:
the chiral currents $W_{s,n}^\pm$ act as derivations of the chiral
ground ring,
and the quantum currents $J$ and $\overline J$ of \ruffox\ are derivations
of the full quantum ground ring.

\section{Determination Of The Ground Ring}

The ground ring is easily determined.  First, we consider the chiral ground
ring.  We note that the product $x^ny^m$ has the quantum numbers
of $\co_{(n+m)/2,(n-m)/2}$, so $\co_{u,n}$
is a multiple of $x^{u+n}y^{u-n}$ if the latter is not zero.
We will prove that $x^ny^m\not= 0$ for all $n,m$ in the next subsection.
Accepting this for the moment, the chiral ground
ring of the $\co_{u,n}$'s is just the
ring of polynomial functions in $x$ and $y$.

Similarly, the anti-chiral ground ring (of BRST invariant antiholomorphic
spin $(0,0)$ operators) consists of polynomial functions in $x'
=\overline\co_{1/2,1/2}$ and $y'=\overline\co_{1/2,-1/2}$.

Now let us combine left and right movers to make the full quantum ground
ring.  It is generated by operators $x^ny^mx'{}^{n'}y'{}^{m'}$ with
$n+m=n'+m'$ (to balance the Liouville momenta).  Such operators are multiplied
in the obvious way, by multiplying the left and right moving parts separately.
At $n+m=0$ there is only
the identity operator.
At $n+m=1$ there are four operators,
$$\eqalign{a_1 & =xx'\cr
           a_2 & =yy'\cr
           a_3 & = xy'\cr
           a_4 & =yx' .\cr}\eqn\piplo$$
The ground ring is generated by the $a_i$ since any monomial
$x^ny^mx'{}^{n'}y'{}^{m'}$ with $n+m=n'+m'$ can obviously be written
as a monomial in the $a_i$.    The $a_i$ obey the
one obvious relation
$$ a_1a_2-a_3a_4=0          \eqn\togo$$
and are otherwise independent.  (In algebraic geometry, one says that
the left hand side of \togo\ is a homogeneous polynomial
whose zeros form
a quadric in ${\bf P}^3$, and the equations \piplo\ are an
isomorphism of this quadric with ${\bf P}^1\times {\bf P}^1$.)

Thus, the ground ring of the theory, at the $SU(2)$ point, is the ring
of polynomial functions on the three dimensional quadric cone $Q$
defined by $a_1a_2-a_3a_4=0$.
In comparing with matrix models, we will learn that the three dimensions
of this cone correspond to the matrix eigenvalue, its canonical momentum,
and the ``time'' of the matrix model Schodinger equation.  At the $SU(2)$
point, there is a complete symmetry among these variables, which is partly
lost upon decompactification.

\section{Determination Of The Chiral Symmetry Algebra}

Now we will determine the symmetry algebra of the theory, by using
the fact that it acts as an algebra of derivations of the ground ring.
First we consider the chiral currents.

To determine how a chiral current $J$ acts on the chiral ground ring,
it is enough, from   \pofgo, to know what it does to $x$ and $y$.
Indeed if
$$J(x)=f(x,y),~~{\rm and}~~J(y)=g(x,y),\eqn\yuggo$$
then it follows from \pofgo\ that for an arbitrary element
$w(x,y)$ of the ground ring, we get
$$J(w)=\left(f{\partial\over\partial x}+g{\partial\over\partial y}\right)
\cdot w.                       \eqn\uggo$$
In particular, $J$ can be identified with the vector field
$f\partial_x+g\partial_y$ on the $(x,y)$ plane, and so the chiral currents
generate an algebra of diffeomorphisms of the $(x,y)$ plane.
(And similarly, the quantum currents which combine left and right
movers act by diffeomorphisms of the quadric cone $Q$.)

Let us determine the vector fields corresponding to some particular
$W_{s,n}^\pm$.  First of all, the $W_{s,n}^-$ are trivially
disposed of.  They have momenta $(n,i(-1-s))\sqrt 2$, and would map
$x$ or $y$ to spin $(0,0)$ operators of ghost number 0 and Liouville momentum $
i(-1/2-s)\sqrt 2$.  As there is no BRST cohomology with those quantum numbers,
the $W_{s,n}^-$ annihilate $x$ and $y$ and therefore the whole ground ring.
Similarly, the cosmological constant operator $W_{0,0}^+$
(which in any case is the same as $W_{0,0}^-$) annihilates $x$ and $y$.
Now we consider the $W_{s,n}^+$.
The first non-trivial case is $W_{1/2,\pm 1/2}^+$.  These
are tachyon operators,
$$W_{1/2,\pm 1/2}^+=e^{(\pm i X+\phi)/\sqrt 2}. \eqn\yurgo$$
Recalling the definition of $y$, we have
$$W_{1/2,1/2}^+(z)\cdot y(w)=e^{(iX+\phi)/\sqrt 2}(z)
\cdot \left(cb-{i\over \sqrt 2}\partial(X+i\phi)\right)\,e^{-(iX+\phi)/\sqrt 2}
(w)\sim {1\over z-w}.        \eqn\xurgo$$
The action of $W_{1/2,1/2}^+$ on $y$ is given by the residue of the pole,
which is 1, so
$$W_{1/2,1/2}^+(y)=1.\eqn\turfogo$$
On the other hand, $W_{1/2,1/2}^+$ annihilates $x$ (there is no short
distance singularity in the operator product $W_{1/2,1/2}^+\cdot x$).
Hence we have determined that $W_{1/2,1/2}^+$ acts on the chiral ground
ring as
$$W_{1/2,1/2}^+={\partial\over\partial y}. \eqn\opurgo$$
Similarly, $W_{1/2,-1/2}^+$ acts as
$$W_{1/2,-1/2}^+={\partial\over\partial x}. \eqn\topurgo$$

At this point, we can fill a gap in our determination of the chiral
ground ring by proving that $x^ny^m\not= 0$.  Indeed, an $n$-fold
repeated commutator of $x^ny^m$ with $W_{1/2,-1/2}^+$ followed by an $m$-fold
repeated commutator with $W_{1/2,1/2}^+$ gives 1, which does not vanish,
so $x^ny^m$ must not be zero.

{}From \opurgo\ and \topurgo, we see that $W_{1/2,1/2}^+$ and $W_{1/2,-1/2}^+$
commute in their action on the $x-y$ plane.  One might be inclined to think
that they commute altogether.  In fact
$$W_{1/2,1/2}^+(z)W_{1/2,-1/2}^+(w)=
e^{(iX+\phi)/\sqrt 2}(z)e^{(-iX+\phi)/\sqrt 2}(w)
\sim {1\over z-w}e^{\phi\sqrt 2}(w)={1\over z-w} W_{0,0}^+(w).\eqn\uppo$$
The object $W_{0,0}^+(w)$ is ``central'' in the sense that it commutes with all
of the currents, as we will see, and acts trivially
on the $x-y$ plane.
\uppo\ is the beginning of the discovery that what is going on in the problem
is quantum mechanics on the $x-y$ plane, as the commutator of (the operators
representing) $\partial_x$ and $\partial_y$ is not zero but is a central
object.

Now, let us determine the rest of the algebra.  The operators
$W_{1,n}^+$ are simply the $SU(2)$ currents.  $x$ and $y$ have
$T_3=\pm 1/2$, respectively (as we see from their momentum values),
and as they are the only cohomology classes with their values of the ghost
number and Liouville momentum, they form an $SU(2)$ representation which
must have spin 1/2.  Thus, without further ado we have
$$\eqalign{W_{1,1}^+ & =x{\partial\over\partial y} \cr
W_{1,0}^+ & ={1\over 2}\left(x{\partial\over\partial x}-y{\partial\over\partial
y}\right) \cr
W_{1,-1}^+ & =y{\partial\over\partial x}. \cr}\eqn\polgoc$$
(These formulas have
been written as if $x$ and $y$ might be real variables, with
symmetry group $SL(2,\IR)$ rather than the $SU(2)$ that is natural in
the Euclidean theory.  Upon complexification, \polgoc\ is equivalent to the
standard $SU(2)$ generators.)

Now we consider the $W_{s,n}^+$ with $s>1$.  These operators transform
under $SU(2)$ with spin $s$.  By considering the Liouville momenta,
we see that they map $x$ or $y$ to polynomials in $x$
and $y$ that are homogeneous of degree $2s-1$.  Such a polynomial
has spin $s-1/2$.  On the other hand, we know that $x$ and $y$ have spin
$1/2$.  In combining spin $s$ with spin $1/2$ to make spin $s-1/2$,
there is only one invariant coupling.  If these couplings are
not zero (which we will
prove presently), they can be set to any desired non-zero value
by scaling the $W_{s,n}^+$ by factors
depending only on $s$.   Therefore, if we can find any Lie algebra
in which these couplings are all non-zero, and in which the $W_{s,n}$
of $s\leq 1$ act correctly, it is the one we want.

\REF\area{I. Bakas, ``The Large $N$ Limit Of Extended Conformal Symmetries,''
Phys. Lett. {\bf B228} (1989) 57, ``Area Preserving Diffeomorphisms And Higher
Spin Fields In Two Dimensions,'' in
{\it Supermembranes And Physics In $2+1$ Dimensions,} ed. M. Duff et. al
(World Scientific, 1990).}
\REF\preserving{C. N. Pope, L. J. Romans, and X. Shen, ``A Brief History
of $W_\infty$,'' in {\it Strings 90}, ed. R. Arnowitt et. al. (World
Scientific, 1991), and references therein.}
In view of suggestions about the possible
relation of area-preserving diffeomorphisms to high spin fields in
two dimensions
[\area,\preserving], and the above observation that $[\partial_x,
\partial_y]$ is central, there is an obvious
Lie algebra with the right properties.
Let $\omega$ be the area form
$$\omega=dx\,\,dy \eqn\huuf$$
on the $x-y$ plane.  Every polynomial $h(x,y)$
determines a vector field
$$ {\partial h\over\partial x}{\partial\over\partial y}-{\partial h\over
\partial y}{\partial\over\partial x} \eqn\uuf$$
that generates an infinitesimal area-preserving transformation
of the $x-y$ plane.  If we identify $W_{s,n}^+$ with the transformation
determined by the vector field
that corresponds to $h(x,y)=x^{s+n}y^{s-n}$, then all of the conditions
are obeyed since (i) these $W$'s are non-zero; (ii) this assignment
is $SU(2)$ (or $SL(2,\IR)$) covariant; (iii) these $W$'s obviously generate
a Lie algebra.  The area preserving diffeomorphisms of the $x-y$ plane
are therefore the Lie algebra that we are looking for.

It remains to show that for any $s$, the $W_{s,n}^+$ act non-trivially
on $x$ and $y$.
First let us note that for any $s$, by counting the Liouville momenta,
the commutator $[W_{1/2,n}^+,
W_{s,n'}^+]$ is a multiple of $W_{s-1/2,n+n'}^+$.
Moreover, for given $s$, these commutators are
not all zero.  Indeed, for $n=-1/2$ and $n'=s$
one has $W_{1/2,-1/2}^+
=e^{(-iX+\phi)/\sqrt 2}$ and $W_{s,s}^+=e^{(isX+(1-s)\phi)\sqrt 2}$.
As
$$e^{(-iX+\phi)/\sqrt 2}(z)\cdot e^{(isX+(1-s)\phi)\sqrt 2}(w)
\sim {1\over z-w}e^{(i(s-1/2)X+(3/2-s)\phi)\sqrt 2}(w)
={1\over z-w}W_{s-1/2,s-1/2}(w), \eqn\hidoc$$
one has
$$ [W_{1/2,-1/2}^+,W_{s,s}^+]=W_{s-1/2,s-1/2}^+.  \eqn\idoc$$
Inductively, if $W_{s-1/2,s-1/2}^+$ does not annihilate $x$ and $y$,
then $W_{s,s}^+$ does not either.  As we know that the $W_{s,n}^+$ do not
all annihilate $x$ and $y$ for sufficiently small $s$, this is true
for all $s$, as assumed above.

It is interesting to note that the ``energy'' operator of the theory,
that is, the operator that generates translations of $X$, is
$\partial X$ which is none other than $W_{1,0}^+$ in our present
notation.
According to the above, this operator generates on the $x-y$ plane
the motion derived from the Hamiltonian function $h(x,y)=xy$.
With $p=x+y$, $q=x-y$,
this is none other than the inverted harmonic oscillator Hamiltonian
$h=p^2-q^2$
familiar in the $c=1$ matrix model.  We will make a more extensive
comparison with the matrix model in \S3,
after combining left and right movers.
The conclusion will be much the same.

\subsection{The Other Operators}

Now we want to extend this discussion to determine the commutators of
the $W_{s,n}^-$.

As $W_{s,n}^-$ has Liouville momentum $-i\sqrt 2(s+1)$,
a commutator $[W_{s,n}^-,W_{s',n'}^-]$ must be a multiple of
$W_{s+s'+1,n+n'}^-$.  But
$W_{s,n}^-$ transforms under $SU(2)$ with spin $s$.  Since spin
$s+s'+1$ does not appear in the tensor product of spin $s$ and spin $s'$,
we must have
$$ [W_{s,n}^-,W_{s',n'}^-]=0, ~~{\rm for \,\,\,all}~s,n,s',n'.\eqn\kopo$$

Next we consider a commutator $[W_{s,n}^+,W_{s',n'}^-]$.
As $W_{s,n}^+$ has Liouville momentum $-i\sqrt 2(-s+1)$,
the Liouville momentum of the commutator
is $-i\sqrt 2(s'-s+2)$.  If $s\geq 1+s'$,
we must write this as $-i\sqrt 2(-s''+1)$ where $s''=s-s'-1$, and
then $[W_{s,n}^+,W_{s',n'}^-]$ is a multiple of $W_{s'',n+n'}^+$.
As the latter transforms under $SU(2)$ with spin $s''=s-s'-1$,
and spin $s-s'-1$ does not appear in the tensor product of spin $s$
with spin $s'$, $[W_{s,n}^+,W_{s',n'}^-]$
necessarily
vanishes for such $s$ and $s'$.  It remains to consider the case $s<1+s'$.
Then we must have
$$[W_{s,n}^+,W_{s',n'}^-]={\rm const}\cdot W_{s'',n+n'}^-, \eqn\hufd$$
with $s''=s'+1-s$.  Therefore the problem is just to determine
how the $W^-$ transform under the algebra of the $W^+$.

I claim that the answer is that (after scaling the $W_{s,n}^-$ by
suitable factors that depend only on $s$), $W_{s,n}^-$ transforms like
$$ W_{s,n}^-\sim {\partial^{(s-n)/2}\over \partial x^{(s-n)/2}}
{\partial^{(s+n)/2}\over\partial y^{(s+n)/2}}\delta^2(x,y). \eqn\ufu$$
I  will just briefly indicate how this can be proved.
The proposed formula is obviously $SU(2)$ covariant, so the
$W_{1,n}^+$ act correctly.  For $s=1/2$, the commutators
$[W_{1/2,1/2}^+,W_{s,s}^-]$ are all non-zero.  This is proved by
explicitly evaluating them; the calculation is just analogous
to the one above for $[W_{1/2,-1/2}^+,W_{s,s}^+]$.
The commutators $[W_{1/2,n}^+,W_{s,n'}^-]\sim W_{s+1/2,n+n'}^-$
are all
determined by $SU(2)$ in terms on one unknown $s$ dependent coefficient, which
is nonzero for all $s$ in view of what was just said.
This being so, the $W_{s,n}^-$ can be uniquely normalized so as to transform
under $W_{1/2,n}^+$ as claimed in \ufu.
The derivatives of $\delta^2(x,y)$ appearing in \ufu\ give a representation
of the algebra of polynomial area-preserving vector fields, so to prove
that the $W_{s,n}^-$ transform as claimed, it is enough to show that
there is only one possible representation in which the $W_{s,n}^+$
for $s\leq 1$ act as claimed.  This is an immediate consequence of
the formula $[W_{1/2,n}^+,W_{s,n'}^+]\sim W_{s-1/2,n+n'}^+$
which combined with $SU(2)$ invariance
uniquely determines
the action of the $W_{s,n}^+$ once the
action of the $W_{s-1/2,n}^+$ is known.

The $W_{s,n}^-$ act trivially on the $x-y$ plane; this can
be demonstrated by considering the Liouville momentum and $SU(2)$
transformations, as above.

\section{The Quantum Symmetry Algebra}

Now we will combine the left and right movers and determine the quantum
symmetry algebra, that is the algebra of the operators introduced earlier
in \ruffox:
$$\eqalign{J_{s,n,n'} & = W_{s,n}^+\overline\co_{s-1,n'} \cr
           \overline J_{s,n,n'} & = \co_{s-1,n}\overline W_{s,n'}^+ \cr}
                       \eqn\uffox$$
I will again proceed indirectly, using the fact that the $J$'s and
$\overline J$'s must correspond to vector fields on the quadric
surface $Q$ defined by
$$ a_1a_2-a_3a_4=0. \eqn\ffox$$

A hypersurface $H$ in affine space, with defining equation
$f=0$, has a natural volume form
$$ \Theta=-{da_1da_2da_3\over \partial f/\partial a_4}.  \eqn\udoo$$
It can be described more invariantly by saying that $\Theta$ is the
unique three form on $H$ such that on $H$
$$\Lambda =-df\cdot \Theta \eqn\oppo$$
where $\Lambda=da_1da_2da_3da_4$ is (up to a constant multiple) the unique
translation and $SL(4)$ invariant volume form on the ambient affine space.
The latter description makes it possible to define $\Theta$
without singling out one of the coordinates.
For the quadric $Q$, $f=a_1a_2-a_3a_4$, so
$$\Theta={da_1da_2da_3\over a_3}. \eqn\nonono$$
If we write
$$a_3=\rho e^\psi,~~a_4=\rho e^{-\psi}, \eqn\opo$$
so that the equation of $Q$ is $\rho^2=a_1a_2$, then
$$\Theta=d\psi da_1 da_2.      \eqn\bopp$$

The polynomial vector fields on $Q$ transform under $SO(2,2)$,
whose complexification is the same as that of $SU(2)\times SU(2)$,
as
$$\bigoplus_{n=0,1/2,1,\dots}\left((n+1,n)\oplus (n,n+1)\oplus (n+1,n+1)\right)
. \eqn\kiffo$$
The volume preserving polynomial vector fields transform as
$$\bigoplus_{n=0,1/2,1,\dots}\left((n+1,n)\oplus (n,n+1)\right).
\eqn\hiffo$$
These facts will be demonstrated in appendix 2.
I claim that the Lie algebra of the $J$'s and $\overline J$'s is
the algebra of volume preserving polynomial vector fields on $Q$,
and in fact that the $J$'s and $\overline J$'s correspond respectively
to the pieces of spin $(n+1,n)$ and $(n,n+1)$.
To demonstrate this, it is enough to note that $J_{s,n,n'}$ and
$\overline J_{s,n,n'}$ have spin $(s,s-1)$ and $(s-1,s)$, for
$s=1,3/2,2,\dots$.  Moreover, the vector fields corresponding to
the $J_{s,n,n'}$ and $\overline J_{s,n,n'}$
are nonzero because of arguments similar to those that we gave in the
chiral case.  The $SU(2)\times SU(2)$ content therefore ensures that the
$J$'s and $\overline J$'s must exactly fill up the volume preserving
polynomial vector fields on $Q$.

\section{Departures From The $SU(2)$ Point}

Now we will briefly discuss how some of the structures change in the
compactified theory under a small departure from the $SU(2)$ point.
Some of the most interesting points will be speculative.

The infinitesimal moduli of the theory, for small departure from the
$SU(2)$ point, are
$$ Z_{s,n,n'}^\pm=W_{s,n}^\pm\overline W_{s,n'}^\pm.  \eqn\tobco$$
I will focus on deformations in which only the $Z^+$ are excited.
(However, the $Z^-$ moduli are also important, for instance in two
dimensional black holes.)
The $Z^+$ moduli transform under $SU(2)\times SU(2)$ as
$$\bigoplus_{n=0,1/2,1,\dots} (n,n).         \eqn\illopo$$
As we will see in appendix (2), this is precisely the $SU(2)\times SU(2)$
content of a general polynomial
function $\phi$ on the hypersurface $Q$.  Thus, the
$Z^+$ moduli naturally combine into such a function.

Now let us try to {\it guess} the geometric meaning of the function
$\phi$.  Such
a function gives precisely the data we need to deform the quadric $Q$
to a general nearby hypersurface
described by an equation
$$a_1a_2-a_3a_4=\phi(a_i). \eqn\llopo$$
I conjecture that the first order $Z^+$ corrections to the operator algebra
deform the relation $a_1a_2-a_3a_4=0$ into \llopo.
In particular, the cosmological constant corresponds to the case $\phi=\mu$
(a constant).  Thus, according to our ansatz, the ground ring at $\mu\not= 0$
but still at the $SU(2)$ radius is the ring of functions on the smooth
quadric
$$ a_1a_2-a_3a_4=\mu.    \eqn\hopo$$

\REF\rabinovici{L.P. Kadanoff and A. C. Brown, ``Correlation Functions On
The Critical Lines Of THe Baxter And Ashkin-Teller Models,''
Ann. Phys. {\bf 121} (1979)
318; R. Dijkgraaf, E. Verlinde, and H. Verlinde, ``$c=1$ Conformal
Field Theories on Riemann Surfaces,'' Comm. Math. Phys.
{\bf 115} (1988) 649; P. Ginsparg, ``Curiosities At $c=1$,''
Nucl. Phys. {\bf B295 [FS]21}
(1988) 153; S. Chaudhuri and J. A. Schwartz, ``A Criterion For
Integrably Marginal Operators,'' Phys. Lett. {\bf 219B} (1989) 291;
A. Giveon, N. Malkin, and E. Rabinovici, ``On Discrete Symmetries And
Fundamental Domains Of Target Space,'' Phys. Lett. {\bf 238B} (1990) 57.}
Thus, conjecturally, to first
order, the hypersurface $Q$ can be deformed to an arbitrary nearby
affine hypersurface in the ambient four-space.  However, the general
deformation will be obstructed in second order by a non-zero beta function.
This can be computed as follows [\rabinovici].  Consider a general
theory with left and right moving currents $J_a$ and $\overline J_a$,
with structure constants $f_{ab}{}^c$ and $\overline f_{ab}{}^c$.  For an
infinitesimal perturbation of the Lagrangian
$$L\to L+\sum w^{aa'}J_a\overline J_{a'}, \eqn\rorf$$
the quadratic beta function corresponds to the counterterm
$$ \beta=w^{aa'}w^{bb'}f_{ab}{}^cf_{a'b'}{}^{c'}J_c\overline J_{c'}.\eqn\orf$$
We can easily implement this prescription in the present situation.
Since the structure constants for left and right movers are just the
Poisson brackets in $x-y$ and $x'-y'$, respectively, the beta function is
$$ \beta=\epsilon^{ij}\epsilon^{i'j'}{\partial^2\phi\over\partial x^i\partial
x'{}^{i'}}{\partial^2\phi\over\partial x^j\partial x'{}^{j'}}.\eqn\jiroof$$
After a relatively long but straightforward calculation, one finds that
this can be written in terms of
the $a_i$ as follows.  Let $Z$ be the scaling operator $Z=\sum_i
a_i\cdot \partial/
\partial a_i$.  Let $\eta_{ij}$ be the metric such that $a_1a_2-a_3a_4=
\eta_{ij}a^ia^j/2$.  Let $\Delta=\eta^{ij}\partial_i\partial_j$
be the corresponding Laplacian.  Then
$$\beta={1\over 2}\left(\eta^{ij}(\partial_iZ\phi)(\partial_jZ\phi)
    -(Z^2\phi)\cdot \Delta \phi\right). \eqn\yiff$$

An important check on this formula is that if $\phi$ is a solution of
the equation $\beta=0$, then so is $\phi'=\phi+(a_1a_2-a_3a_4)f$ for any $f$;
indeed
we wish to identify $\phi$ and $\phi'$, as they coincide on $Q$.
As will be explained in the appendix, there is always a unique choice
of $f$ such that $\Delta \phi'=0$.
It is easy to see that if $\phi$ is a homogenous polynomial
(so $Z\phi=\lambda\phi$ for some $\lambda$)  such that $\Delta \phi=0$
and $\phi$ is a solution of $\beta=0$, then $w(\phi)$ is also
a solution of $\beta=0$ for any function $w$.  The simplest nonconstant
function
obeying the stated conditions is $\phi=(a_1a_2+a_3a_4)/2$.
Consequently (since in fact $\phi=a_1a_2$ on $Q$)
$$\overline\phi=-\sum_{n=0}^\infty\epsilon_n(a_1a_2)^n\eqn\popolo$$
is a perturbation with vanishing second order beta function, for any
values of the constants $\epsilon_n$.

According to our ansatz for the meaning of $\phi$, the marginal operator
\popolo\ corresponds to deforming the
quadric $Q$ to a more general hypersurface
$$a_3a_4=a_1a_2+\sum_{n=0}^\infty\epsilon_n(a_1a_2)^n. \eqn\joppo$$
In comparing to matrix models, as we will see,
the eigenvalue phase space will correspond
to the $a_1-a_2$ plane,
and the curve $a_3a_4=0$ in the $a_1-a_2$ plane will correspond to
the fermi surface.  The function $a_1a_2$ will correspond to the standard
inverted harmonic oscillator Hamiltonian $H=p^2-q^2$, and the function on
the right hand side of \joppo\ is the Hamiltonian of a perturbed matrix model.

The problem can probably be analyzed more completely using a criterion of
Chaudhuri and J. A. Schwarz, who claim [\rabinovici] that a perturbation
of the type \rorf\ is exactly marginal if and only if the $w^{aa'}$
are non-zero only for $a$ and $a'$ in some abelian subalgebra of the
full current algebra.

Actually, under a generic perturbation of the compactified theory
near the $SU(2)$ point, much of the structure we have found disappears.
For instance, at a generic value of the radius of the $X$ field,
it is necessary to match the left and right moving momenta, and so to discard
most of the states.  This instability of the structure seems not to have
an analog in the uncompactified theory.  Its proper interpretation is unclear.

\chapter{The Uncompactified Theory; Comparison To The Matrix Model}

We now want to consider the
{\it uncompactified} string theory.  Henceforth, therefore, $X$ will
be real valued, rather than circle valued.  Of course, this is the case
that is best understood from the point of view of matrix models.

Of the operators that we have found in
\S2, the ones that survive in the uncompactified theory are the ones
that have $p_L-p_R=0$, where $p_L$ and $p_R$ are the left and right moving
momenta.
Recalling the definitions
$$\eqalign{a_1 & = xx' \cr a_2 & = yy' \cr a_3 & = xy' \cr a_4 & = yx' \cr}
              \eqn\ddoof$$
and noting that the $(p_L,p_R)$ values are
$$\eqalign{ x:&~~(1,0)/\sqrt 2 \cr
            y:&~~(-1,0)/\sqrt 2 \cr
            x':&~~(0,1)/\sqrt 2 \cr
            y':&~~(0,-1)/\sqrt 2 \cr} \eqn\wwoof$$
we see that the allowed operators are precisely the ones that are
invariant under
$$K=a_3{\partial\over\partial a_3}-a_4{\partial\over\partial a_4}. \eqn\woofo$$
If we set $a_3=\rho e^\psi,~~a_4=\rho e^{-\psi}$ as in \S2.5, then
$$K={\partial\over\partial \psi}.           \eqn\tugomo$$
Any $K$ invariant polynomial in the $a_i$ is in fact a polynomial in $a_1$
and $a_2$. (The product $a_3a_4$ is invariant, but equals $a_1a_2$.)
Any symmetry of the model must therefore induce a motion of the $a_1-a_2$
plane -- which will turn out to be the eigenvalue phase space of the matrix
model.

Now, let us determine the symmetry group of the uncompactified model.
This is simply the subgroup of the group of volume-preserving diffeomorphisms
of the quadric cone $Q$ which are also $K$ invariant.
We recall from \bopp\ that the volume form in these coordinates is
$$\Theta=d\psi\,\,da_1\,\,da_2. \eqn\ugomo$$
A general $K$ invariant vector field is of the form
$$ f(a_1,a_2){\partial\over\partial a_1}+g(a_1,a_2){\partial\over\partial a_2}
          +u(a_1,a_2){\partial\over\partial\psi}.  \eqn\gomo$$
Requiring that this preserve $\Theta$, we see that there is no restriction on
$u$, while
$$f{\partial\over\partial a_1}+g{\partial\over\partial a_2} \eqn\omo$$
must generate a diffeomorphism of the $a_1-a_2$ plane that preserves
the area form $da_1da_2$.  Thus area-preserving diffeomorphisms of the
$a_1-a_2$ plane appear in the uncompactified theory.  The operators
$$u(a_1,a_2){\partial\over\partial \psi} \eqn\tomo$$
commute with each other, and transform under diffeomorphisms of the
$a_1-a_2$ plane as abelian gauge transformations.  The combined structure
is thus ``gravity'' (area-preserving diffeomorphisms) plus abelian
gauge theory on the $a_1-a_2$ plane.

There is an important gap in this discussion,
however.  While every volume preserving
$K$ invariant vector field on the quadric $Q$ determines
a vector field
\gomo\ as just described, the converse is not true.  We have to determine
what happens to $a_3$ and $a_4$ when a given transformation of $\psi-a_1-a_2$
space is ``lifted'' to a transformation of the quadric.
As $a_3=\sqrt{a_1a_2}e^\psi,~a_4=\sqrt{a_1a_2}e^{-\psi}$, one immediately sees
that
$$\eqalign{\left(f{\partial\over\partial a_1}+g{\partial\over\partial a_2}
          +u{\partial\over\partial \psi}\right)&a_3={a_3\over 2}\cdot\left(
 {f\over a_1}+{g\over a_2}\right)+u a_3 \cr
\left(f{\partial\over\partial a_1}+g{\partial\over\partial a_2}
          +u{\partial\over\partial \psi}\right)&a_4={a_4\over 2}\cdot\left(
 {f\over a_1}+{g\over a_2}\right)+u a_4 .\cr}\eqn\hhh$$
This is a polynomial vector field in the space of the $a_i$ only if
$f$ is divisible by $a_1$ and $g$ is divisible by $a_2$.

The latter requirement means that the symmetries of the theory include
not all of the area preserving polynomial vector fields on the $a_1-a_2$
plane, but {\it only those that preserve the locus $a_1a_2=0$.}
I claim that the $a_1-a_2$ plane corresponds to the eigenvalue
phase space of the matrix model.  $a_1$ and $a_2$ correspond to
$p+q$ and $p-q$, where $q$ is the matrix eigenvalue and
$p$ is its canonical momentum.  The locus $a_1a_2=0$ corresponds to
{\it the fermi surface of the matrix model},
that is, it is the surface on which the inverted harmonic oscillator
Hamiltonian $H=p^2-q^2$ vanishes.
When we study the matrix model presently, we will see why area preserving
diffeormorphisms that preserve the fermi surface arise naturally.

Intuitively, the distinguished role of the locus $a_1a_2=0$ comes
about because this locus is somewhat analogous to the ``branch points''
of the projection from the quadric $Q$ to the $a_1-a_2$ plane.
The defining equation $a_3a_4=a_1a_2$ of the quadric shows that the
subspace of $Q$ lying over a generic point in the $a_1-a_2$ plane is
a hyperbola,
while over a point with $a_1a_2=0$ one has a pair of lines.  Because
of this difference,
a diffeomorphism of the $a_1-a_2$ plane that does not map the locus
$a_1a_2=0$ to itself cannot be ``lifted'' to a $K$ invariant diffeomorphism
of $Q$.

In \S2.6, I have described a hopefully
very plausible ansatz according to which
general $K$ invariant perturbations of the theory will perturb
the defining equation of $Q$ to
$$ a_3a_4=h(a_1,a_2), \eqn\todoco$$
with $h$ an arbitrary Hamiltonian function. If so,  then the ``branch locus''
would be the curve $h=0$ in the $a_1-a_2$ plane.  $K$ invariant volume
preserving transformations of the three-manifold \todoco\ correspond to
area preserving diffeomorphisms of the $a_1-a_2$ plane that map the locus
$h=0$ to itself, plus the abelian gauge transformations $u\cdot \partial/
\partial\psi$.  This would be expected if $h$ corresponds to the  one
body Hamiltonian of the matrix model.
In particular, according to the ansatz, when the cosmological constant
is turned on,  the hypersurface
$a_1a_2-a_3a_4=0$ is deformed to a hypersurface $Q_\mu$ given by
$a_1a_2-a_3a_4=\mu$.  A $K$ invariant volume
preserving diffeomorphism of $Q_\mu$ is an area preserving diffeomorphism
of the $a_1-a_2$ plane that leaves fixed the curve $a_1a_2-\mu=0$.  This
curve corresponds to the fermi surface of the matrix model at chemical
potential $\mu$.

\section{Symmetries Of The Matrix Model}

\REF\avan{J. Avan and A. Jevicki, ``Classical Integrability And
Higher Symmetries of Collective String Field Theory,''BROWN-HET-801 (1991).}
\REF\polch{D. Minic, J. Polchinski, and Zhu Yang, ``Translation-Invariant
Backgrounds in $1+1$ Dimensional String Theory,'' UTTG-16-91.}
Let us now analyze the symmetries of the matrix model version of the
$c=1$ theory.  Our discussion will be fairly similar to that of [\moore,\avan,
\polch],
but we will consider a somewhat larger class of symmetries.  The extra
symmetries might be considered ``trivial'' in the matrix model but are
important in comparing to the conformal field theory.

Consider the Lagrangian for a free particle moving in the $p-q$ plane:
$$L=\int dt\left(p\dot q - H(p,q)\right).\eqn\filo$$
This can be written
$$L=\int \alpha, \eqn\ilo$$
where $\alpha$ is the one form
$$\alpha =p\,\,dq-H\,\,dt. \eqn\olo$$
A symmetry generator is an arbitrary infinitesimal transformation
$$\eqalign{\delta p & = f(p,q,t) \cr
           \delta q & = g(p,q,t) \cr
           \delta t & = u(p,q,t) \cr}
\eqn\jolo$$
that leaves the Lagrangian invariant.  To this end, it is not necessary
to leave $\alpha$ invariant.  A transformation under which $\alpha\to
\alpha+d\beta$ is also a symmetry.  Instead of the one form $\alpha$, we should
consider the two form $\omega=d\alpha=dp\,\,dq-dH\,\,dt$.  A symmetry
is a transformation that leaves $\omega$ invariant.

Let us consider the matrix model with the standard inverted harmonic
oscillator Hamiltonian, $H(p,q)=(p^2-q^2)/2$.  Then $\omega=dp\,\,dq
-(p\,\,dp-q\,\,dq)dt.$  If we set
$$\eqalign{p' & = p \cosh t - q\sinh t \cr
           q' & = -p \sinh t +q\cosh t \cr} \eqn\tolo$$
then we find
$$\omega =dp'\,\,\,dq'.            \eqn\poolo$$
It is therefore easy to identify the vector fields that preserve
$\omega$:
they are of the form
$$\left({\partial g(p',q')\over \partial q'}{\partial\over\partial p'}
-{\partial g(p',q')\over \partial p'}{\partial\over\partial q'}\right)
+u(p',q,t){\partial\over\partial t}.            \eqn\opolo$$
Note that $g$ is a function of $p',q'$ only but $u$ may also depend
on $t$.
This is very similar to the answer that we found in the conformal
field theory, with $a_1,a_2\leftrightarrow p',q'$ and
$\psi\leftrightarrow t$.
There are two discrepancies: (i) in the conformal field
theory $u$ was a function of $a_1$ and $a_2$ only, and not $\psi$; (ii)
in the conformal field theory
$g$ was required to obey a certain restriction, such that the surface
$a_1a_2=0$ is invariant under the symmetry generated by $g$.

I cannot explain the first discrepancy, but there is a very nice way
to parametrize it.  Let $\Theta$ be the volume form in $(p,q,t)$ space
$$\Theta=dp\,\,dq\,\,dt=dp'\,\,dq'\,\,dt. \eqn\dopolo$$
The transformation in \opolo\ preserves $\Theta$ as well as $\omega$
precisely if $u$ is a function of $p',q'$ only.
The invariances of the conformal field theory are thus the transformations
that
leave invariant the volume form $\Theta$ as well as the Lagrangian
\ilo\ and the surface $a_1a_2=0$.

As for that last restriction, which is our second apparent discrepancy,
it has a very nice explanation in
the matrix model.
The $(1,0)$ and $(0,1)$ currents that we found
in the conformal field theory are unbroken symmetries of a particular
ground state.  They
should be compared to the symmetries of the matrix model that {\it preserve
the ground state}.  The ground state of the matrix model is
the state in which all the single particle levels
of $H(p,q)<0$ are filled and the others are empty.   An area preserving
transformation that maps filled states to filled states and empty states
to empty states necessarily maps the fermi surface to itself.
As $H(p,q)=(p^2-q^2)/2=(p'{}^2-q'{}^2)/2$, the area preserving transformations
of the $p'-q'$ plane that leave fixed the matrix model ground state
are those that leave fixed the fermi surface $p'{}^2-q'{}^2=0$.
(It must be left fixed as a set, not pointwise.)  The fermi surface
is obviously isomorphic to the fixed locus $a_1a_2=0$ of the conformal
field theory, with $a_1=p'+q'$, $a_2=p'-q'$.

Thus, we find the dictionary for comparing the conformal field theory
to the matrix model: $(a_1,a_2,\psi)$ corresponds to $(p'+q',
p'-q',t)$.

Let us give an example  of the use of this dictionary.
The difference between the left and right moving
$X$ momenta, which corresponds to the difference $\partial X-\bar\partial X$,
annihilates all states in the (uncompactified) conformal field theory.
It is mapped under our correspondence to
$$\left.{\partial\over\partial t}\right|_{p',q'}, \eqn\boo$$
which annihilates all matrix model observables since the single particle
equation of motion of the matrix model
equation of motion is $dp'/dt=dq'/dt =0$.
The time translation operator of the conformal field theory corresponds
to the current $\partial X+\bar\partial X$.  This is mapped to the
canonical transformation of the $p'-q'$ plane generated by the
vector field derived from the Hamiltonian $a_1a_2=p'{}^2-q'{}^2$ -- the
usual inverted harmonic oscillator, in other words.

We have carried out this discussion at zero cosmological constant,
but if the ansatz in \S2.6 is correct, then the generalization to
nonzero cosmological constant -- and arbitrary perturbation of the matrix
model Hamiltonian -- is immediate.

\subsection{Conjectured Interpretation}

We will conclude by briefly suggesting a partial
interpretation of some of these results.
The symmetries of the matrix model preserve both a two form $\omega$
and a three form $\Theta$.  Since $\omega$ depends explicitly on the
matrix model Hamiltonian, it seems to be a dynamical variable.  On the
other hand, $\Theta$ may well be more universal.
Near the $SU(2)$ point, there seems to be always a volume form analogous
to $\Theta$ but not necessarily a two form $\omega$.  One is tempted
to try to think of the theory as a three dimensional theory, with
a symmetry group of volume preserving diffeomorphisms, in which one of
the fields (analogous to the metric tensor in general relativity)
is a two form $\omega$ whose equations of motion assert that $d\omega=0$.
The symmetries of a given classical solution would be the subgroup of the
group of volume preserving diffeomorphism that leaves fixed both
$\omega$ and the other degrees of freedom, such as the tachyon field.
It would be a challenge to reconcile such a viewpoint with the rich
structure that exists upon compactification, for instance near the
$SU(2)$ point.

\ack{I am indebted to R. H. Dijkgraaf, D. J. Gross, A. M. Polyakov,
E. Rabinovici, and N. Seiberg for many stimulating discussions.}

I understand that I. Klebanov and A. M. Polyakov
have recently succeeded in computing directly the current algebra
of the chiral spin one discrete fields.

\endpage
\centerline{APPENDIX (1): BRST Analysis Of Some Low-Lying States}

\def\dw{\downarrow}
\def\uw{\uparrow}
\def\rg{\rangle}
\def\lt{\left}

Our goal in this appendix is to describe the lowest non-trivial
examples of some phenomena analyzed theoretically in [\zuckerman].
First we recall the standard construction of the BRST operator.
It is
$$Q=\sum_nc_nL_{-n}-{1\over 2}\sum_{n,r}c_{-n}c_{-r}b_{n+r}(n-r), \eqn\offo$$
where $c_n, \,b_m,\,\,\,n,m\in \IZ$ are the Fourier modes of ghosts
and antighosts, which obey $\{c_n,b_m\}=\delta_{n+m}$, and $L_n$ are
the Virasoro generators of matter.  The latter are
$$L_n={1\over 2}\sum_{n\in \IZ}\left(\alpha_m\alpha_{n-m}
     +\phi_m\phi_{n-m}\right)+i\sqrt 2 n\phi_n,\eqn\ffo$$
with $\alpha_n$ and $\phi_n$ (obeying $[\alpha_n,\alpha_m]=
[\phi_n,\phi_m]=n\delta_{n+m}$) being the Fourier modes of $X$ and of the
Liouville field $\phi$.
In particular, $\alpha_0$ and $\phi_0$ are
the $X$ and $\phi$ momentum operators.
\foot{In
going from states, we will consider in this appendix, to operators,
there is a shift $\phi_0\to\phi_0-i\sqrt 2$.  The Liouville momentum
values given in the text were the values for the operators.}
The Fock vacuum will be labeled $|p,\uparrow\rangle$
or $|p,\downarrow\rangle$, where $p=(\alpha_0,\phi_0)$ labels
the momentum eigenvalues,
$|p,\downarrow\rangle$ is a state annihilated by $b_0$, and
$|p,\uw\rg=c_0|p,\dw\rg$.

The above expressions for the $L$'s and $Q$ should all
be normal ordered.  The normal ordering
constants are zero for this two dimensional system (the ``tachyon''
is a massless particle in $D=2$).

We will define the ghost number operator, which we will call $G$, such that
$|p,\dw\rg$ has $G=0$ (from some points of view $G=-1/2$ is more natural).

Interesting phenomena occur
at special values of $\alpha_0$ and $\phi_0$.  We will
consider the first two levels.
In addition to the points (involving cohomology at $G=-1$) that were important
in this paper, there are other peculiarities of the BRST cohomology that
may be important in the future.  To exhibit these peculiarities, I will
describe the level one situation in full detail.  When we get to level
two, we will just focus on $G=-1$.  The other peculiarities of level one
are repeated at level two (and at all higher levels, according to
[\zuckerman]), but the detailed description would be somewhat long.

\section{Level One}

At level one, discrete states arise at $\alpha_0=0$, $\phi_0=\pm i\sqrt 2$.
We will write the formulas for a more generic situation with
arbitrary $\alpha_0$ and $(\alpha_0{}^2+\phi_0{}^2)/2=-1$; this condition
ensures that $L_0$ annihilates all the states that we consider.
We lose nothing by that restriction, since it is easy to prove that the
BRST cohomology arises only at $L_0=0$.
We will consider the states in order of ascending values of the
ghost number.

At level one and $G=-1$, there is only one state,
$b_{-1}|p,\downarrow\rangle$, with
$$Qb_{-1}|p,\downarrow\rangle=\left(\alpha_0\alpha_{-1}+(\phi_0-i\sqrt 2)
        \phi_{-1}\right)|p,\downarrow\rangle.\eqn\mimodd$$

For $G=0$, there are three states, with
$$\eqalign{Qb_{-1}|p,\uw\rg & =\lt(\alpha_0\alpha_{-1}+(\phi_0-i\sqrt 2)
            \phi_{-1}\right)|p,\uw\rg -2c_{-1}|p,\dw\rg\cr
           Q\alpha_{-1}|p,\dw\rg & = \alpha_0c_{-1}|p,\dw\rg \cr
           Q\phi_{-1}|p,\dw\rg & = \left(\phi_0+i\sqrt 2\right)c_{-1}|p,\dw
       \rg .\cr}\eqn\turfo$$

For $G=1$, we get
$$\eqalign{ Q\phi_{-1}|p,\uw\rg & = \left(\phi_0+i\sqrt 2\right)c_{-1}
          |p,\uw \rg \cr
            Q\alpha_{-1}|p,\uw\rg & = \alpha_0c_{-1}|p,\uw\rg \cr
            Qc_{-1}|p,\dw\rg & = 0 .\cr} \eqn\gurfo$$

Finally, at $G=2$,
$$ Qc_{-1}|p,\uw\rg = 0 . \eqn\jurfo$$

It is easy to see that the $Q$ cohomology is trivial in this space
at generic $\alpha_0$.  Something interesting happens at $\alpha_0=0$.
At that value of $\alpha_0$, the $L_0$ condition permits two values of
$\phi_0$, namely $\pm i\sqrt 2$.  The behavior at those two  values are
completely different.  First we consider the value $\phi_0=i\sqrt 2$
(which is shifted to $\phi_0=0$ when one goes from states to operators).

In this case there is a one dimensional cohomology at $G=-1$, since
$Qb_{-1}|p,\dw\rg=0$ for such $p$.  This state corresponds to the identity
operator 1, which is the first of the spin zero, ghost number zero
operators that were the main focus of the body of the paper.
($G$ as we have defined it is increased by one in going from states
to operators, so states of $G=-1$ will correspond to operators of $G=0$,
and so on.)
At $G=0$, the cohomology is actually {\it two} dimensional.
For representatives we can take $\alpha_{-1}|p,\dw\rg$ and
$\phi_{-1}|p,\dw\rg +i\sqrt 2b_{-1}|p,\uw\rg$.
The former corresponds to the spin zero, $G=1$ operator
$c\partial X$ and to the spin one, $G=0$ operator $\partial X$.
The latter is the first of the ghost number zero discrete currents
constructed from the bosons only (without the ghosts) and usually
considered in discussions of the discrete states.  The latter
corresponds to the spin zero, $G=1$ operator $c\partial\phi+\sqrt 2
\partial c$ but does not correspond to any spin one, $G=0$ current.
The reason for this will be explained presently.
Finally, to complete the enumeration of the cohomology at this value
of the momenta, the $G=1$ state $\alpha_{-1}|p,\uw\rangle$
corresponds to the $G=2$ spin zero operator $c\partial c\partial X$,
but does not correspond to any current.

\subsection{Relation Between Spin Zero Operators And Currents}

Let us now briefly explain why certain of the states just found
do not correspond to currents.
First of all, given a BRST cohomology class, to find a corresponding
BRST invariant spin zero primary field we should find a representative
$|\psi\rg$ of the cohomology class that is of Virasoro highest weight,
obeying $L_n|\psi\rg=0,\,\,\,n\geq 0$.  In the cases at hand this
is possible; the representatives given above satisfy this
condition.  In particular, as the highest weight of these states is zero,
they correspond to primaries of spin zero.  To get a current, we need
a highest weight state of spin one.  The general strategy for doing this
is to set
$$|\alpha\rangle =b_{-1}|\psi\rangle.    \eqn\momo$$
If this vanishes, we cannot proceed further; this is why the $G=-1$
state $b_{-1}|p,\dw\rg$ does not correspond to a current.
Suppose $|\alpha\rg\not= 0$.
Then
$$L_0|\alpha\rg=L_0b_{-1}|\psi\rg=b_{-1}|\psi\rg=|\alpha\rg, \eqn\jxo$$
so if $|\alpha\rg$ is a highest weight state, the highest weight is
one, and the operator corresponding to $|\alpha\rg$ will be a current.
Moreover,
$$Q|\alpha\rg=Qb_{-1}|\psi\rg =L_{-1}|\psi\rg.\eqn\msoo$$
Although this is not zero, it will correspond to a total
derivative (since acting on operators $L_{-1}\sim \partial/\partial z$);
this is good enough.  The question that remains
is whether $|\alpha\rangle$ is a highest weight vector, that is, whether
$L_n|\alpha\rangle$ is zero, or at least a BRST commutator, for $n>0$.
We have
$$L_n|\alpha\rg =L_nb_{-1}|\psi\rg =b_{n-1}|\psi\rg. \eqn\ripop$$
Now these states may not vanish, but we do at least have
$$ Qb_{n-1}|\psi\rg =L_{n-1}|\psi\rg =0,\,\,\,n>0, \eqn\ipop$$
as $|\psi\rg$ is BRST invariant and highest weight.  Therefore,
if the BRST invariant states $b_{n-1}|\psi\rg$ are BRST commutators,
then $|\alpha\rg$ is of highest weight, at least modulo BRST commutators.

For two of the cohomology classes found above, this last condition
fails.  The obstructions arise for $n=1$.  For $|\psi\rangle
=\phi_{-1}|p,\dw\rg+i\sqrt 2 b_{-1}|p,\uw\rg$, we get $b_0|\psi\rg
=-i\sqrt 2 b_{-1}|p,\dw\rg$, which is not a BRST commutator; indeed
it represents the now familiar $G=-1$ cohomology class.  For $|\psi\rangle
=\alpha_{-1}|p,\uw\rg$, we get $b_0|\psi\rangle=\alpha_{-1}|p,\dw\rg$,
the standard discrete state at $G=0$.

Actually, the only possible obstruction is the one at $n=1$ because
$L_{n-1}\|\psi\rangle$ is an $L_0$ eigenstate with eigenvalue
$n-1$, and hence automatically a BRST commutator unless $n=1$.

\subsection{The Other Liouville Dressing}

Now we consider what happens for the other Liouville dressing,
$\alpha_0=0$, $\phi_0=-i\sqrt 2$.  The important point for the
present paper is that nothing interesting happens at $G=-1$, so
we do not get any new operators
of spin zero and $G=0$.  At $G=0$, we have $\alpha_{-1}|p,\dw\rg$,
which corresponds to the spin zero operator $c\partial X e^{2\sqrt 2\phi}$
and to the current $\partial X\,\,e^{2\sqrt 2\phi}$; the latter,
when combined with right movers, becomes the marginal operator associated
with the black hole.  At $G=1$, the cohomology is two dimensional.
One state is $\alpha_{-1}|p,\uw\rangle$,
which corresponds to the spin zero, $G=2$ operator $c\partial c\partial X
e^{2\sqrt 2\phi}$ but not to any spin one current (because of the obstruction
described above).  The second is  $\phi_{-1}|p,\uw\rg$,
which corresponds to a spin zero operator $c\partial c\partial \phi \,\,
e^{2\sqrt 2\phi}$ of $G=2$ and to a current $\partial c\partial\phi \,\,
e^{2\sqrt 2\phi}$ of $G=1$
that
is a spin one conformal field up to a BRST commutator.
Finally, at $G=2$ we have $c_{-1}|p,\uw\rg$, which corresponds to the
spin zero $G=3$ field $c\partial c\partial^2c \,\,\,e^{2\sqrt 2\phi}$, but not
to any current.

\section{Level Two}

At level two, the mass shell condition, which we may as well impose,
is $(\alpha_0{}^2+\phi_0{}^2)/2=-2$.  The space of states at $G=-1$
is three dimensional, generated by $b_{-2}|p,\dw\rg$, $b_{-1}\alpha_{-1}
|p,\dw\rg$, and $b_{-1}\phi_{-1}|p,\dw\rg$. We get
$$\eqalign{
  Qb_{-2}|p,\dw\rg & =\left(\alpha_0\alpha_{-2}+{1\over 2}\alpha_{-1}^2
         +\phi_0\phi_{-2}+{1\over 2}\phi_{-1}^2-2i\sqrt 2\phi_{-2}
                   +3c_{-1}b_{-1}\right)
                         |p,\dw\rg                         \cr
  Qb_{-1}\alpha_{-1}|p,\dw\rg &= \left(
            \alpha_0c_{-1}b_{-1} +\alpha_0\alpha_{-1}^2+\alpha_{-2}+
              (\phi_0-i\sqrt 2)\phi_{-1}\alpha_{-1}\right)|p,\dw\rg \cr
  Qb_{-1}\phi_{-1}|p,\dw\rg &  = \left(\alpha_0\alpha_{-1}\phi_{-1}
       +(\phi_0-i\sqrt 2)\phi_{-1}^2+\phi_{-2}+(\phi_0+i\sqrt 2)c_{-1}b_{-1}
              \right)|p,\dw\rg
         .\cr} \eqn\ududu$$

The kernel of $Q$ in this space vanishes unless $\alpha_0^2=1/2$, that is,
$\alpha_0=\epsilon/\sqrt 2$ with $\epsilon=\pm 1$.  There are then two possible
Liouville dressings, with $\phi_0=\pm 3i/\sqrt 2$.  At $\phi_0=-3i/\sqrt 2$,
$Q$ remains acyclic at $G=-1$, as the reader can easily verify from
\ududu.  But at $\phi_0=3i/\sqrt 2$, the cohomology is one dimensional,
generated by
$$\left(b_{-2}-{\epsilon\over\sqrt 2}b_{-1}\alpha_{-1}+{i\over \sqrt 2}
        b_{-1}\phi_{-1}\right)|p,\dw\rg. \eqn\toxo$$
These states correspond to the spin zero, $G=0$
operators
$$\eqalign{ x & = \left(cb+{i\over \sqrt 2}(\partial X-i\partial\phi)\right)
                \cdot e^{i(X+i\phi)/\sqrt 2} \cr
y & = \left(cb-{i\over \sqrt 2}(\partial X+i\partial\phi)\right)
                \cdot e^{-i(X-i\phi)/\sqrt 2} \cr} \eqn\oxoxo$$
that were discussed in the text and shown to generate multiplicatively
the entire ring of spin zero, $G=0$ operators.
They also correspond to the spin one, $G=-1$ currents
$b\cdot \exp({\pm i(X\pm i\phi)/\sqrt 2})$.

The other phenomena that we found at level one all recur at level two.
In the text, the important states at $G\geq 0$ were $G=0$ states
that give rise to currents.  At level two, these are
$$\left(\alpha_{-2}-\epsilon\sqrt 2\alpha_{-1}^2\right)|p,\dw\rg,
          \eqn\ipobu$$
with either Liouville dressing.
The corresponding currents are
$$W_{3/2,\epsilon/2}^\pm=\left(\partial^2X-i\epsilon\sqrt 2
(\partial X)^2\right)\,\,\exp(i\epsilon X/\sqrt 2+\phi(2\mp 3)/\sqrt 2).
\eqn\nugo$$

\endpage
\centerline{APPENDIX (2): Functions And Vector Fields On Certain Quadrics}

In section two we required a knowledge of how polynomial
functions and vector
fields on the quadric $a_1a_2-a_3a_4=0$ transform under $SO(2,2)\sim
SL(2,\IR)\times
SL(2,\IR)$.
As they transform in finite dimensional representations, and the classification
of such representations is discrete, nothing can change if we make
a continuous perturbation
to the quadric $a_1a_2-a_3a_4=\mu$ (which in any case is physically
relevant according to the ansatz in \S2.6).  Moreover, as the discussion
is just a question of algebra, we could just as well change the
signature and consider the standard sphere
$$\sum_{i=1}^4 x_i^2 = 1      \eqn\momo$$
in $\IR^4$.

The $x_i$ transform under $SU(2)\times SU(2)$ in the $(1/2,1/2)$
representation.  We can make this explicit by writing them as $x^{AA'}$,
$A,A'=1,2$; the $A$ and $A'$ indices transform as $(1/2,0)$ and
$(0,1/2)$, respectively.
An $n^{th}$ order polynomial
$$\phi=w_{A_1A_2\dots A_n;A_1'A_2'\dots A_n'}x^{A_1A_1'}x^{A_2A_2'}
\dots x^{A_nA_n'}, \eqn\momodo$$
with $w$ symmetric in all $A_i$ and all $A'_j$, transforms as spin $(n/2,n/2)$.
Any polynomial function
on the sphere is a linear combination of these, for the following reason.
If one were to antisymmetrize in a pair of indices, say $A_1$ and $A_2$,
to make $\epsilon_{A_1A_2}x^{A_1A'_1}x^{A_2A'_2}$, then by bose statistics
one may as well antisymmetrize in $A'_1$ and $A'_2$.
But $\epsilon_{A_1A_2}\epsilon_{A'_1A'_2}x^{A_1A'_1}x^{A_2A'_2}=\sum_i
x_i{}^2$.  Such a factor can be discarded, as it equals 1 on the sphere.
This justifies the claim made in \S2 that the polynomial functions
on the sphere transform as
$$\bigoplus_{n=0,1/2,1,\dots}(n,n). \eqn\oido$$

Regarded as a function on $\IR^4$ (dropping the restriction to
$x_ix^i=1$), $\phi$ is a solution of
the Laplace equation $\Delta\phi=0$ (where $\Delta=\sum_i\partial^2/\partial
x_i^2$) for the following reason.  $\Delta\phi$ is a polynomial in the
$x$'s of degree $n-2$ that must transform with spin $(n/2,n/2)$; as $x$
has spin $(1/2,1/2)$, this is impossible.
This justifies the claim in \S2.6 that every polynomial on the sphere
can be extended on $\IR^4$ as a solution of the Laplace equation.
Conversely, any polynomial on $\IR^4$ that obeys the Laplace equation
is a linear combination of the $\phi$'s.  To show this, it is enough,
by the $SU(2)\times SU(2)$ symmetry and the homogeneity of the Laplace
operator, to show that for $n>0$,
$\left(\sum_ix_i^2\right)^n\cdot (x^{11'})^m$ is not a solution of
the Laplace
equation.  This is easy if one recalls that the Laplacian is
$\Delta=\epsilon^{AB}\epsilon^{A'B'}\partial_{AA'}\partial_{BB'}$.

Any polynomial vector field on $\IR^4$ is of the form
$$\sum_{i=1}^4f^i(x^j){\partial\over\partial x^i} \eqn\umpo$$
with each $f^i$ being a function.
The $\partial/\partial x_i$ transform as $(1/2,1/2)$.  For
each $i$, the space of possible $f^i$'s, modulo functions that
vanish on the sphere, transforms as $\bigoplus_{n\geq 0}(n,n)$.
By combining $(1/2,1/2)$ with $\bigoplus_{n\geq 0}(n,n)$, we
get
$$\bigoplus_{n\geq 0}(n,n)\bigoplus_{m\geq 1/2}(m,m)\bigoplus_{r\geq 0}
\left((r+1,r)\oplus (r,r+1)\right). \eqn\humbo$$
We want to restrict this to vector fields on the sphere, that is vector
fields that leave the surface $\sum_ix_i^2=1$ fixed, or in other
words vector fields such that $x_if^i=0$ on the sphere.
In general, $x_if^i$ would be an arbitrary function on the sphere,
transforming as \oido, so imposing the equation $x_if^i=0$ amounts
to dropping the first term in \humbo.  Thus, the $SU(2)\times SU(2)$
content of the polyomial vector fields on the sphere is actually
$$\bigoplus_{m\geq 1/2}(m,m)\bigoplus_{r\geq 0}
\left((r+1,r)\oplus (r,r+1)\right).\eqn\jumobo$$

If we want volume preserving vector fields, we must take the orthocomplement
of the vector fields of the type
$$ f^i=\partial^iw          \eqn\umobo$$
for any function $w$.  The $w$'s would transform like
\oido, but a constant term in $w$ is irrelevant.  Removing the vector
fields \umobo\ thus amounts to removing the first term in \jumobo, and
thus the volume preserving polynomial vector fields on the sphere
transform under $SU(2)\times SU(2)$ as
$$\bigoplus_{r\geq 0}
\left((r+1,r)\oplus (r,r+1)\right)\eqn\moboo$$
as was claimed in \S2.  The various pieces can be described very explicitly.
The vector fields
$$w_{A_1,\dots,A_{n};A'_1\dots A'_{n+2}}
x^{A_1A'_1}\dots x^{A_{n}A'_{n}}x^{BA'_{n+1}}
\epsilon^{A'_{n+2}B'}
      {\partial\over\partial x^{BB'}}
\eqn\yolo$$
transform as spin $(n/2,n/2+1)$.  Volume preserving vector fields
of spin $(n/2+1,n/2)$ are constructed similarly.

\refout
\end
\bye